\documentclass[useAMS,usenatbib,letterpaper]{mn2e}

\usepackage{txfonts,graphicx,amssymb,rotating}

\title[Fragmenting discs with short cooling times]{Binary formation and mass function variations in fragmenting discs with short cooling times}

\author[Alexander et al.]
  {R.D.Alexander$^{1,2}$\thanks{email: rda@strw.leidenuniv.nl}, P.J.Armitage$^{2,3}$ 
  and J.Cuadra$^{2}$\\
  $^1$ Sterrewacht Leiden, Universiteit Leiden, Niels Bohrweg 2, 2300 RA, Leiden, the Netherlands\\
  $^2$ JILA, University of Colorado, Boulder, CO 80309-0440, USA\\
  $^3$ Department of Astrophysical and Planetary Sciences, University of Colorado, Boulder, CO 80309-0391, USA}

\begin{document}

\pagerange{\pageref{firstpage}--\pageref{lastpage}} \pubyear{2008}

\maketitle

\label{firstpage}

\begin{abstract}
Accretion discs at sub-pc distances around supermassive black holes are likely to cool rapidly enough that self-gravity results in fragmentation. Here, we use high-resolution hydrodynamic simulations of a simplified disc model to study how the outcome of fragmentation depends upon numerical resolution and cooling time, and to investigate the incidence of binary formation within fragmenting discs.  We investigate a range of cooling times, from the relatively long cooling time-scales that are marginally unstable to fragmentation down to highly unstable cooling on a time-scale that is shorter than the local dynamical time. The characteristic mass of fragments decreases with reduced cooling time, though the effect is modest and dependent upon details of how rapidly bound clumps radiate. We observe a high incidence of capture binaries, though we are unable to determine their final orbits or probability of survival. The results suggest that faster cooling in the parent disc results in an increased binary fraction, and that a high primordial binary fraction may result from disc fragmentation.  We discuss our results in terms of the young massive stars close to the Galactic Centre, and suggest that observations of some stellar binaries close to the Galactic Centre remain consistent with formation in a fragmenting accretion disc. 
\end{abstract}


%%%%%%%%%%%%%%%%%%%%%%%%%%%

\section{Introduction}\label{sec:intro}
In recent years the topic of gravitational instabilities in accretion discs has been the subject of a great deal of theoretical and numerical research.  Such gravitational instabilities have applications both to accretion disc theory, in terms of the transport of angular momentum, and also to theories of star and planet formation, as under certain circumstances the instability can lead to fragmentation \citep[e.g.][]{durisen_ppv,lodato08}.  In particular, the fragmentation of accretion discs has been invoked to explain both the formation of gas giant planets \citep[e.g.][]{boss97,mayer02}, and also to explain the formation of young, massive stars observed close to super-massive black holes at the centres of galaxies \citep[e.g.,][]{ps77,lb03}.

In order for a gravitationally unstable disc to fragment, two conditions must be met.  The first is that the disc must be sufficiently cold, or sufficiently massive, to satisfy the \citet{toomre64} criterion
\begin{equation}\label{eq:Q}
Q = \frac{c_{\mathrm s} \kappa}{\pi G \Sigma} \lesssim 1
\end{equation}
Here $c_{\mathrm s}$ is the local sound speed, $\kappa$ the epicyclic frequency (which in a thin disc is approximately equal to the local orbital frequency), and $\Sigma$ the disc surface density.  In addition, in order to fragment the disc must cool sufficiently rapidly to overcome compressional heating during collapse.  \citet{gammie01} used a local analysis, and found the fragmentation boundary to be
\begin{equation}
t_{\mathrm {cool}} < \beta_{\mathrm {crit}} \Omega^{-1} \, ,
\end{equation}
where $t_{\mathrm {cool}}$ is the local cooling time-scale, $\Omega$ is the orbital frequency, and $\beta_{\mathrm {crit}} \simeq 3$.  Subsequent studies using global simulations have verified the validity of this criterion \citep{rice03,rice05,meija05,boley07}, although the exact value of $\beta_{\mathrm {crit}}$ required for fragmentation can be larger by a factor of 2--3 depending on the details of the thermal physics adopted (cooling law, equation of state, etc.).  To date, however, most work has either made use of simple thermal physics \citep[e.g.][]{rice03,rice05,nayak07}, or used more realistic thermal physics in the specific case of self-gravitating protoplanetary discs \citep[e.g.][]{cai08,sw08}.  More generally, most studies of fragmentation have tended to look at cases of critical cooling (with $\Omega t_{\mathrm {cool}} \simeq \beta_{\mathrm {crit}}$), despite the fact that some discs, notably those around black holes, are thought to cool much more rapidly \citep[e.g.][]{bl01,goodman03}.

In this paper we use numerical hydrodynamics to study the influence of rapid cooling on the fragmentation properties of gravitationally unstable accretion discs.  In Section \ref{sec:method} we present the details of our numerical simulations, in which we adopt a cooling law consistent with radiative cooling in an optically thick disc.  We use this model to study the non-linear evolution of the instability over a wide range in cooling times.  We also consider in detail the effects of numerical resolution on our calculations, varying the number of fluid elements by a factor of 64 between our lowest- and highest-resolution calculations.  We present our results in Section \ref{sec:results}, first considering issues of numerical convergence and then addressing the effects of varying the cooling time.  In Section \ref{sec:dis} we discuss the limitations of our analysis, and the implications of our results for the formation of stars near to the Galactic Centre, before summarizing our conclusions in Section \ref{sec:summary}.

%%%%%%%%%%%%%%%%%%%%%%%%%%
\section{Numerical Method}\label{sec:method}
The simulations presented in this paper use the publicly-available smoothed-particle hydrodynamics (SPH) code {\sc gadget2} \citep{springel05}.  We have modified the code to include a simple cooling prescription (discussed below), which is representative of the cooling expected in a real accretion disc.   We adopt the standard \citet{mg83} prescription for the SPH viscosity (with $\alpha_{\mathrm {SPH}}=1.0$), using the ``Balsara-switch'' \citep{balsara95} to limit the artificial shear viscosity (as specified in Equations 11--12 of \citealt{springel05}).  We allow for a variable gravitational softening length, and fix the SPH smoothing and gravitational softening lengths to be equal throughout.  We use the standard Barnes-Hut formalism to compute the gravitational force tree \citep[as described in][]{springel05}, and use $N_{\mathrm {ngb}}=64\pm2$ as the number of SPH neighbours.  Typically, a handful of SPH particles end up in rather close orbits around the central gravitating mass and, if left unchecked, result in unreasonably short time-steps.  Consequently we use a single sink particle for the central gravitating mass.  The sink particle simply accretes all gas particles that pass within its radius \citep[as described in][]{cuadra06}, and we set the sink radius to be 1/4 of the inner disk radius.  This is simply a numerical convenience adopted in order to save computing time, and has no physical effect on the simulations.  The simulations were run on the {\sc tungsten} Xeon Linux cluster at NCSA\footnote{See {\tt http://www.ncsa.uiuc.edu/}}, using 64 parallel CPUs for the highest-resolution runs.

Numerical simulations of gas disc dynamics suffer from a basic computational limitation, namely that the pressure scale-height of the disc acts as a minimum length-scale that must always be well-resolved.  Requirements on computing time are therefore very strongly dependent on the disc aspect ratio (the required time typically scales as $(H/R)^{-3}$ or steeper), and accurate global simulations of thin discs are very computationally expensive.  In a self-gravitating disc with $Q\simeq1$, the disc aspect ratio is approximately equal to the ratio of the disc mass to the central object mass, and to date most numerical simulations have focussed on relatively massive, thick discs (with  $M_{\mathrm {disc}} \sim 0.1 M_*$, e.g., \citealt{rice03,rice05,boley07}).  This is typical of young protoplanetary discs, but discs around black holes may in fact be much thinner ($H/R<10^{-2}$).  We therefore seek a method of simulating the behaviour expected in a very thin disc, but without incurring the severe computational penalty that usually results. 

We make use of an artificial disc model, which behaves like a thin self-gravitating disc while in fact remaining thick enough to be well-resolved even at moderate resolution.  Spiral density waves in self-gravitating discs tend to be dominated by a relatively narrow range in spiral mode number, with progressively thinner discs resulting in ever higher-order spiral waves \citep[see the review by][]{lodato08}.  Moreover, as the disc thickness is reduced, the ``local approximation'' (that global modes of the instability are negligible) becomes more accurate.  Consequently, we adopt a relatively large disc mass, but impose an artificial radial cooling profile that results in the disc fragmenting in a single, near-circular filament.  The analysis of our simulations considers only this filament, with the rest of the disc regarded as a boundary condition: the dynamics of fragmentation in a single, near-circular filament are a good approximation to a real self-gravitating disc in the thin disc limit.  With this set-up we are able to resolve the disc scale-height properly with only moderate numbers of SPH particles (see Section \ref{sec:sims} below), and can thus run simulations with much larger particle numbers (``over-resolving'' the disc thickness) to study the fragmentation process in detail when the cooling time is short.

\subsection{Initial Conditions}\label{sec:ics}
We adopt initial conditions similar to those that have been adopted in previous studies of disc fragmentation \citep[e.g.][]{rice03,rice05,rda08}.  We adopt a surface density profile
\begin{equation}
\Sigma(r) \propto r^{-1}
\end{equation}
and a disk temperature that scales as 
\begin{equation}
T(r) \propto r^{-1/2} \, ,
\end{equation}
where $r$ is the (cylindrical) radius.  The disc sound speed $c_s \propto T^{1/2}$ is normalized so that the Toomre parameter (Equation \ref{eq:Q}) is equal to 2.0 at the outer boundary.  The disc is initially vertically isothermal, so the vertical structure of the disc is given by
\begin{equation}
\rho(r,z) = \frac{\Sigma(r)}{\sqrt{2\pi}H} \exp\left(-\frac{z^2}{2H^2}\right) \, ,
\end{equation}
where $H=c_{\mathrm s}/\Omega$ is the disc scale-height (here $\Omega=\sqrt{GM_*/r^3}$ is the Keplerian orbital frequency.).  With this set-up the initial disc is marginally gravitationally stable, and the instability is allowed to develop in a physical manner as the disc cools.  We define a disc mass of $M_{\mathrm d} = 0.1M_*$ (where $M_*$ is the mass of the central gravitating mass), and model the disc using $N_{\mathrm {SPH}}$ SPH particles of mass $m = M_{\mathrm d}/N_{\mathrm {SPH}}$.  We adopt a system of units where $M_*=1$, the inner edge of the disc is at $r=1$, and the time unit is the orbital period at $r=1$.  (Consequently, $G=4\pi^2$ in code units.)  The initial distribution of the SPH particles is obtained by randomly sampling the disc density profile described above, with each particle given a circular velocity in the $x-y$ plane.  We adopt an adiabatic equation of state throughout, with adiabatic index $\gamma=7/5$ (as expected for a disc consisting primarily of molecular hydrogen).

\subsection{Cooling}\label{sec:cooling}
The principle aim of this study is to investigate the effect of variable cooling rates on the fragmentation properties.  At each time-step, we allow the internal energy of the $i$th particle, $u_i$, to cool as
\begin{equation}
\left(\frac{du}{dt}\right)_{\mathrm {cool},i} = -\frac{u_i}{t_{\mathrm {cool}}} \, .
\end{equation}
The cooling time-scale $t_{\mathrm {cool}}$ is defined in terms of the dimensionless parameter $\beta$ and the orbital frequency
\begin{equation}
t_{\mathrm {cool}} = \frac{\beta}{\Omega} \, .
\end{equation}
As mentioned in Section \ref{sec:intro}, previous studies have found that the fragmentation boundary lies in the range $\beta_{\mathrm {crit}}\simeq3$--7.5, with a weak dependence on the adopted equation of state \citep[e.g.][]{gammie01,rice03,rice05}.  Here, in order to study the effects of variable cooling, we define $\beta$ to be a function of radius, density and time:
\begin{equation}
\beta  = {\cal R}(r)  D(\rho)  T(t) \, .
\end{equation}
As discussed above, the radial dependence of $\beta$ is chosen to mimic the behaviour of a tightly-wound spiral density pattern \citep[as expected in a thin disc; e.g.][]{lodato08}, but without the excessive demands on computing resources that usually arise when simulating very thin discs.  We choose $\cal R$ so that the disc fragments in a ring at $r=2$, and can therefore save computing power by considering a limited range in radius.  The only requirement on the radial range is that the boundaries of the disc do not influence the fragmentation, so we fix the inner boundary at $r=1$ and the outer boundary at $r=3$.  The radial dependence of the cooling time is chosen to be a Gaussian profile centred on $r_0=2$
\begin{equation}
{\cal R}(r) = 8.0 \times \left[1 - \frac{1}{2} \exp\left( - \frac{(r-r_0)^2}{2 \Delta ^2} \right) \right] \, ,
\end{equation}
with $\Delta = 0.25$.  Thus ${\cal R}(r)$ varies from 4--8, with the minimum value at $r=r_0$.  The normalisation parameters are chosen so that the disc is initially marginally stable to fragmentation everywhere except close to $r=r_0$, and the value of $\Delta$ is chosen so that the radial length-scale over which the cooling time varies is a few times larger than the pressure scale-height of the disc.

In addition, we allow the cooling time to vary with the local density.  In a real self-gravitating disc we expect radiative cooling to be optically thick (see the discussion in Section \ref{sec:dis}), so the cooling rate is density-dependent.   If cooling is optically thick then the radiative cooling time-scale is proportional to the photon diffusion time-scale.  In the case of constant opacity $\kappa$, the time, $t_{\mathrm D}$, for a photon to diffuse a length $L$ is given by
\begin{equation}
t_{\mathrm D} = \frac{L^2}{c}\rho \kappa \, ,
\end{equation}
where $c$ is the speed of light.  In a collapsing (thinning) disc the appropriate length-scale is the disc thickness $h$, and the density $\rho \simeq \Sigma/h$.  Thus, for 2-D collapse we have
\begin{equation}
t_{\mathrm D} \propto h \propto \rho^{-1} \, ,
\end{equation}
and we see that the cooling time-scale decreases with increasing density $\rho$.  By contrast, in a collapsing spherical clump of mass $m$ and radius $h$, we have $\rho \simeq m/h^3$, so
\begin{equation}
t_{\mathrm D} \propto h^{-1} \propto \rho^{1/3} \, ,
\end{equation}
and therefore the cooling time-scale increases (i.e.~cooling becomes less efficient) as the clump becomes more dense.  This increase in the cooling time-scale can lead to significant suppression of cooling as the forming fragments reach high density, and consequently we adopt a density-dependence with the following form:
\begin{equation}
D(\rho) = \mathrm {max}\left[ 1 \, , \, \left(\frac{\rho}{\rho_0}\right)^{1/3} \right] \, .
\end{equation}
Thus for densities below $\rho_0$ the cooling time is independent of the density; at higher densities the cooling time increases as $\rho^{1/3}$.  Empirically, we choose $\rho_0 = 1.0$ in code units (central object mass / inner disc radius$^3$), as this is typical of the local density in our simulations at the point where bound, roughly spherical objects begin to form.  In practice the weak dependence of the cooling time on density means that our results do not depend significantly on the exact value of $\rho_0$.

Lastly, in order to study the effect of very rapid cooling we allow the cooling rate to vary with time in the simulations.  The motivation for this is that one cannot simply begin a numerical simulation with a cooling time that is much shorter than the dynamical time-scale, as the resulting collapse and fragmentation merely amplifies noise in the initial conditions and has little or no physical meaning.  Instead, one must allow the instability to develop in a physical manner before the cooling time is allowed to become short.  We follow the approach of \citet{clarke07}, and allow the cooling time to decrease as
\begin{equation}
T(t) = 1 - \left(\frac{1}{\tau}\frac{t}{t_{\mathrm{orb}}(r_0)}\right)
\end{equation}
Here $t_{\mathrm {orb}}$ is the local orbital period, and $\tau$ is a dimensionless parameter that we use to parametrize the cooling rate.  We note that the form of $T(t)$ allows the cooling time to become very small (and indeed negative) at large $t$; in practice, however, our simulation run for short enough times that this is not a problem.  We adopt values of the cooling parameter $\tau = 2$, 3, 5, $\infty$ (in the latter case the cooling time is constant with time).  With this set-up the disc is marginally unstable to fragmentation at $r=2$ for $\tau = \infty$, while smaller values of $\tau$ result in cooling times that are up to order of magnitude shorter.  At the end of our simulations the smallest values of $\beta$ are in the range 0.2--0.3 (see also Table \ref{tab:sims}).

\subsection{Simulations}\label{sec:sims}
\begin{table}
 \centering
  \begin{tabular}{cccc}
  \hline
$N_{\mathrm {SPH}}$ & $\tau$ & $t_{\mathrm {frag}}$ & $\beta'(t_{\mathrm {frag}})$ \\
  \hline
200,000 & $\infty$ & 8.5 & 4.0\\
1,600,000 & $\infty$ & 8.2 & 4.0\\
12,800,000 & $\infty$ & 8.1 & 4.0\\
200,000 & 5 & 6.6 & 2.1\\
1,600,000 & 5 & 6.4 & 2.2\\
200,000 & 3 & 5.7 & 1.3\\
1,600,000 & 3 & 5.7 & 1.3\\
200,000 & 2 & 5.0 & 0.5\\
1,600,000 & 2 & 4.8 & 0.6 \\
12,800,000 & 2 & 4.8 & 0.6\\
\hline
\end{tabular}
  \caption[List of simulations]{List of simulations run, showing resolution and physical parameters for each simulation.  $t_{\mathrm {frag}}$ denotes the time (in code units) when the first fragment forms, and $\beta'(t_{\mathrm {frag}})$ indicates the approximate value of the local cooling parameter $\beta'(t_{\mathrm {frag}})  = {\cal R}(r_0) T(t_{\mathrm {frag}})$ (i.e.~neglecting the density dependence) when fragmentation occurs.}\label{tab:sims}
\end{table}

The issue of numerical resolution is critical when considering calculations of fragmentation, as insufficient resolution can result in spurious suppression or amplification of fragmentation.  We have conducted a number of calculations using a range of numbers of SPH particles for different values of $\tau$ (see Table \ref{tab:sims}).  Resolution requirements for SPH simulations of fragmentation were presented and tested by \citet{bb97}, and extended for the specific case of self-gravitating discs by \citet{nelson06}.  In this case the Jeans mass must be resolved into $\gtrsim 10 N_{\mathrm {ngb}}$ particles, and the disc scale-height $H$ must be resolved into at least 4 SPH smoothing lengths in the saturated state of the instability\footnote{Note that the definition of the SPH smoothing length in \citet{springel05} differs from that in most of the SPH literature, being larger by a factor of 2.  Throughout this paper we discuss only the standard definition of the smoothing length \citep[e.g.][]{monaghan92}.}.   For most of the models considered here the second of these is the more stringent condition, but correctly resolving the Jeans mass becomes more demanding in the simulations with $\tau=2$ (see Section \ref{sec:res_conv}).  We choose our lowest resolution so that we marginally satisfy these conditions for the case $\tau = \infty$, and thus choose $N_{\mathrm {SPH}} = 200,000$.  We then move to higher resolution, first increasing the particle number by a factor of 8 (to $N_{\mathrm {SPH}} = 1.6\times10^6$), and then by a further factor of 8 (to $N_{\mathrm {SPH}}=1.28\times10^7$).  In a perfectly spherical configuration increasing $N_{\mathrm {SPH}}$ by a factor of 8 results in a factor of 2 ($=8^{1/3}$) increase in the spatial resolution.  However, in the disc geometry considered here we find that a factor of 8 increase in $N_{\mathrm {SPH}}$  in fact improves the spatial resolution by a factor of approximately 2.5.  For convenience, we subsequently refer to the simulations with $N_{\mathrm {SPH}} = 200,000$ as ``low resolution'' (LR), $N_{\mathrm {SPH}} = 1.6\times10^6$ as ``medium resolution'' (MR), and $N_{\mathrm {SPH}}=1.28\times10^7$ as ``high resolution'' (HR).  For reasons of computation time we ran the HR simulations only for the extreme cases $\tau=\infty$ and $\tau=2$; the LR and MR runs simulations were conducted across the full range in $\tau$.

An additional benefit of moving to resolutions higher than the minimum requirement is the potential to resolve binary and multiple systems.  Roughly speaking, we expect that individual objects will only be able to fragment into multiple systems if they cool on a time-scale shorter than the local dynamical time-scale.  In a self-gravitating disc with $Q\simeq 1$ the dynamical time-scale of a single fragment is, to first order, equal to the local orbital time-scale.  Consequently, we do not expect the formation of ``fission'' binaries in discs with $\beta >1$ (although binaries may still form by dynamical capture events), as thermal pressure acts to smooth out density fluctuations on a time-scale shorter than the cooling time-scale.  Thus the minimum resolution requirement is sufficient in simulations with $\beta > 1$.  However, with faster cooling it may be possible to form binary and multiple systems via the dynamical fission of single clumps.  Consequently our high-resolution simulations act not only as convergence tests, but also provide a means of studying the small-scale fragmentation that may occur in discs with short cooling times.

In order to compare the fragmentation in different simulations we must first identify individual fragments (``clumps'') in our simulations, and we do this simply by adopting a density threshold of $\rho=10.0$ (i.e.~ten times larger than the density above which the cooling time begins to slow).  We identify clumps as objects which have peak densities above this threshold, and estimate clump masses by moving outward in spherical shells one (minimum) SPH smoothing length in radius, until the mean density drops below the threshold value: all particles within this radius are then deemed to be part of the clump.  

One consequence of using a fixed density threshold in this manner is that clumps are identified at somewhat earlier times in simulations with higher resolution, as the central regions of individual clumps are no longer ``smeared out'' over as long a length-scale.  This is an artefact of our analysis, rather than a defect in the simulations, and we solve this simply by comparing simulations at different resolutions relative to the time at which the first clump is identified.  In practice this difference is very short: the delay in $t$ between the identification of the first fragment at the highest and lowest resolutions is a few percent of the total simulation time (see Table \ref{tab:sims}).  Overall this is a rather crude method for measuring the clump mass function, and a more sophisticated approach could, for example, allow for non-spherical clumps, or test every particle in order to determine how many individual particles are bound to each density peak \citep[e.g.][]{rda08}.  In practice, however, such ``improvements'' create at least as many problems as they solve, because the treatment of, for example, binary and multiple systems, or ``circumstellar'' discs, becomes ambiguous.  The method used here is robust across different resolutions, and is sufficiently accurate for our needs: {\it a posteriori} checks show that negligible fraction of clumps identified in this manner are not in fact bound objects

%%%%%%%%%%%%%%%%%%%%%%%%%%%%%
\section{Results}\label{sec:results}
\subsection{Numerical convergence}\label{sec:res_conv}
\begin{figure}
\centering
        \resizebox{\hsize}{!}{
        \includegraphics[angle=270]{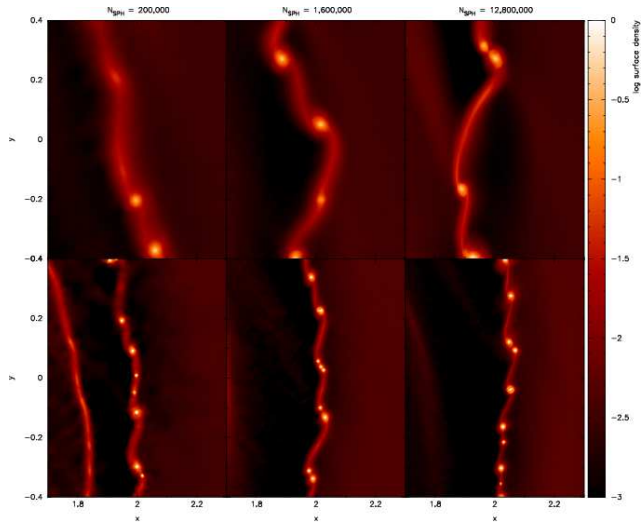}
        }
        \caption{Plots of the disc surface density in the simulations with the slowest ($\tau=\infty$; top row) and fastest ($\tau = 2$; bottom row) cooling times, using different numerical resolutions.  From left-to-right, the three columns show the LR, MR and HR simulations respectively, and each plot is made 1/8 of a local orbital period (0.35 time units) after the first fragment forms.  For slow cooling, just below the fragmentation threshold, we see that the fragmentation process is well-resolved in all our simulations, although more substructure is apparent in the higher-resolution simulations.  With fast cooling the disc is colder when it fragments, and higher resolution is needed in order to resolve the disc scale-height: in this case we see that the LR simulation is somewhat under-resolved (e.g.~ the density peaks at $r\simeq1.8$ are numerical artefacts), but that $N_{\mathrm {SPH}} = 1.6\times10^6$ appears sufficient to resolve the fragmentation process properly.}
        \label{fig:conv}
\end{figure}
We first seek to compare our results at difference numerical resolutions, in order to test convergence in our simulations.  The non-linear development of gravitational instabilities essentially amplifies seed fluctuations in the density field, and it is not possible to construct identical noise fields at different particle numbers.  Consequently we instead seek statistical convergence, measuring when the first bound objects appear, how many such objects we find, and how much mass is present in these objects.  We identify clumps using the method described in Section \ref{sec:sims}, and use the mass functions derived in this manner to compare different simulations.

\begin{figure}
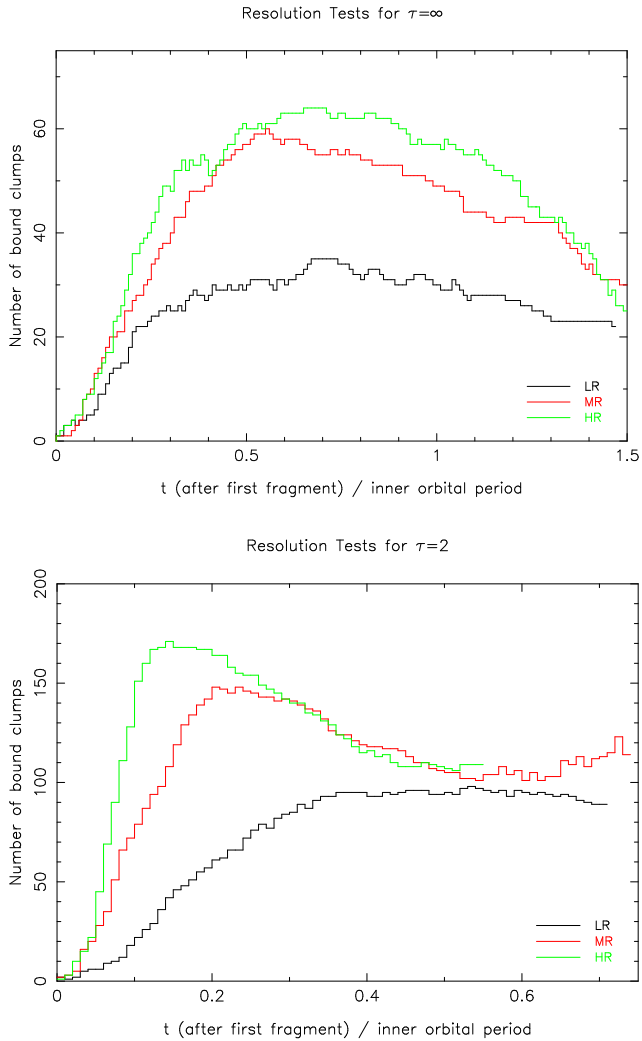

\centering
       \resizebox{\hsize}{!}{
       \includegraphics[angle=270]{fig2a.ps}
	}
	
	\vspace*{5mm}
       
       \resizebox{\hsize}{!}{
       \includegraphics[angle=270]{fig2b.ps}
       }
       \caption{Results of numerical convergence tests.  The upper panel shows the evolution of the number of clumps for the simulations with $\tau=\infty$.  The origin of the time axis is the time when the first clump is identified, and the LR, MR and HR simulations are denoted by black, red and green lines respectively.  Although more binaries are resolved at higher resolution the total mass in clumps does not vary significantly between the runs, indicating a good degree of numerical convergence.  The lower panel shows the same plot for the simulations with $\tau=2$.  In this case the LR simulation is rather under-resolved, but good convergence is found between the MR and HR simulations.}     
           \label{fig:conv_num}
\end{figure}

The results of our numerical convergence tests are shows in Figs.\ref{fig:conv} and \ref{fig:conv_num}\footnote{Visualisations of our SPH simulations were created using {\sc splash}: see \citet{splash} for details.}.  In the case of slow cooling ($\tau=\infty$) we see that the fragmentation process is well-resolved in all of our simulations.  In terms of the criteria laid out by \citet{bb97} and \citet{nelson06}, we resolve the disc scale-height at the mid-plane into approximately 5.3 SPH smoothing lengths in the LR run, 12.5 smoothing lengths in the MR run, and 29.6 smoothing lengths in the HR run.  In addition the characteristic fragment mass is well-resolved throughout so, as noted in Section \ref{sec:sims}, the LR simulation marginally satisfies the minimum resolution requirement.  Runs with higher resolution reveal more substructure, resolving binaries and ``circumstellar'' discs, but  fragmentation is independent of resolution.  Subsequent interactions between clumps, however, are not independent of resolution: this is clearly seen in Fig.\ref{fig:conv_num}, where the number of clumps beyond $\Delta t\simeq0.3$ differs significantly between the LR and HR runs. This is due to mergers of binary clumps, and indicates that the interactions between such binaries are not properly resolved.  This occurs because although the global disc is well-resolved the many ``circumstellar'' discs are not, so the time-scale on which binary pairs merge is governed by the rate of numerical dissipation of energy in these ``sub-discs''.  Consequently we do not attach great significance to our results beyond the time at which binaries start to merge.  Before clump mergers start to become significant the total mass in clumps varies by less than 20\% between the LR  and HR runs, indicating that good numerical convergence has been achieved.

In the case of fast cooling ($\tau=2$) the disc is colder when it fragments, so higher resolution is needed in order to resolve the Jeans mass.  Here, the pressure scale-height of the disc is again marginally resolved in the LR run (into 4.5 SPH smoothing lengths at the disc mid-plane), but the characteristic fragment mass is $\simeq 4N_{\mathrm {ngb}}m$.  Thus in this case, as expected, we see that the LR simulation is slightly under-resolved (by a factor of approximately 2 in particle number), but both resolution requirements are satisfied comfortably in the MR and HR runs.  Binary mergers are somewhat more significant in the MR simulation than in the HR simulation (again due to under-resolving ``circumstellar'' discs), but otherwise there is good agreement between the MR and HR runs.  In both the $\tau=\infty$ and $\tau=2$ cases good statistical convergence is found between the MR and HR runs, so we conclude that $N_{\mathrm {SPH}} = 1.6\times10^6$ is sufficient to resolve the fragmentation process properly throughout all our simulations.

\subsection{Cooling}\label{sec:res_cool}
\begin{figure*}
\centering
        \resizebox{\hsize}{!}{
        \includegraphics[angle=270]{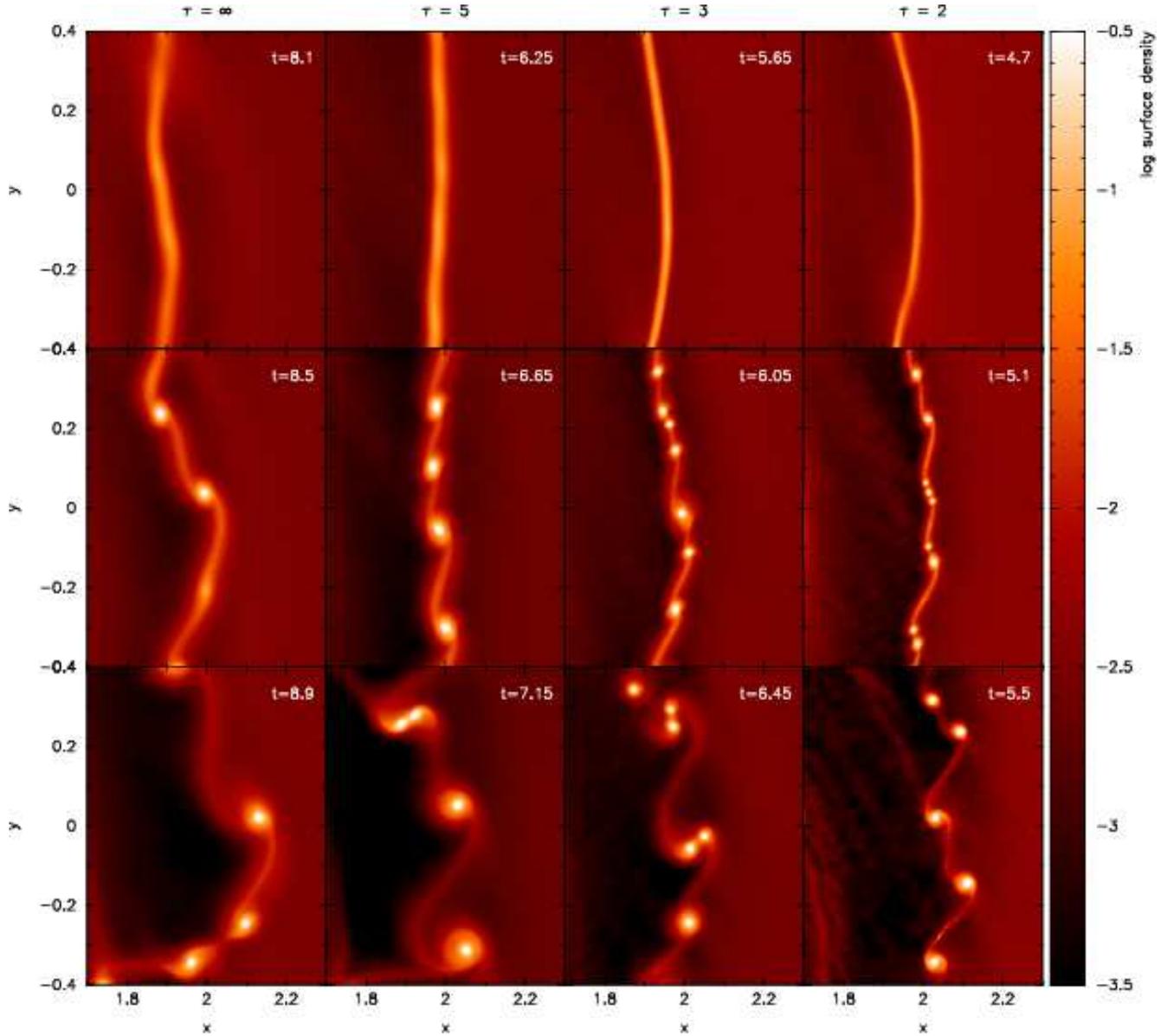}
        }
        \caption{Plots of the disc surface density in the MR simulations, showing how fragmentation develops as a function of cooling time.  From left-to-right, the four columns show the evolution of the simulations with $\tau = \infty$, 5, 3 and 2 respectively.  The plots are made in a reference frame that is co-rotating with the disc at $r=2$, so that only relative motions are evident in the time evolution.  (In this frame the disc shear flow appears to move upwards for $x<2$ and downwards for $x>2$.)  It is clear from the snapshots that faster cooling results in the formation of more, smaller fragments, as well as a higher incidence of multiple systems.  Note also the large number of binary mergers seen between the second and third snapshots in the simulations with fast cooling.}
        \label{fig:panels}
\end{figure*}
\begin{figure}
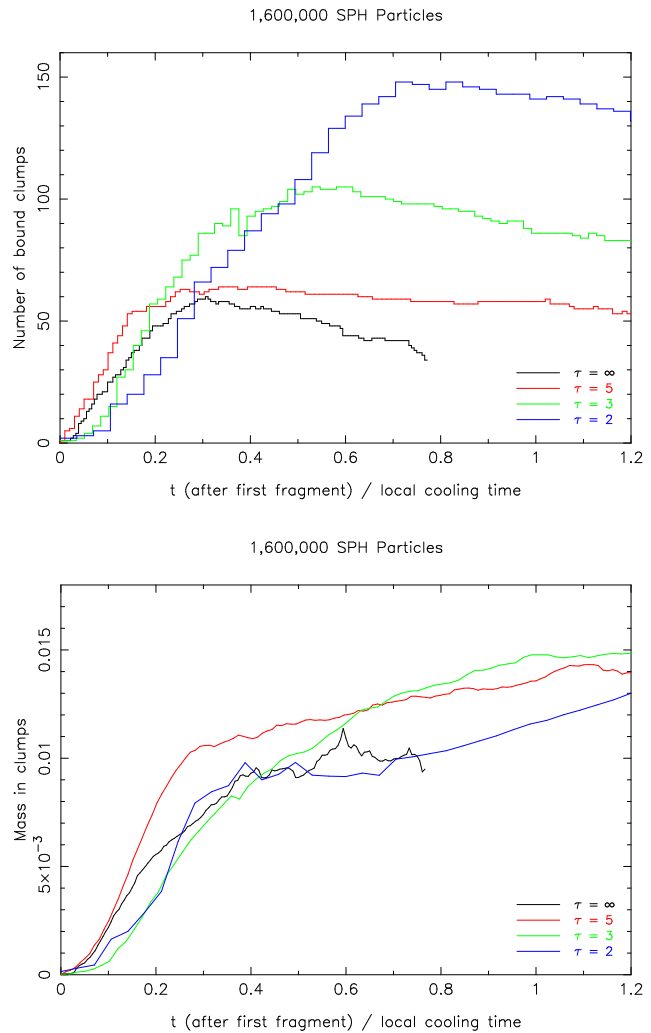

\centering
        \resizebox{\hsize}{!}{
        \includegraphics[angle=270]{fig4a.ps}
        }

	\vspace*{5mm}
	
        \resizebox{\hsize}{!}{
        \includegraphics[angle=270]{fig4b.ps}
        }
        \caption{Fragmentation as a function of cooling time.  The upper panel shows the number of clumps as a function of time in each of the four MR simulations, while the lower panel shows the total mass in the clumps.  The origin of the time axis is set at the point where the first fragment forms, and time is plotted in units of the local cooling time (i.e.~$\beta'(t_{\mathrm{frag}})/\Omega(r_0)$, with the values of $\beta'(t_{\mathrm {frag}})$ taken from Table \ref{tab:sims}).  The total mass in clumps is independent of the cooling time, but we see that faster cooling results in the formation of more objects (with lower masses).}
        \label{fig:cooling}
\end{figure}
\begin{figure}
\centering
        \resizebox{\hsize}{!}{
        \includegraphics[angle=270]{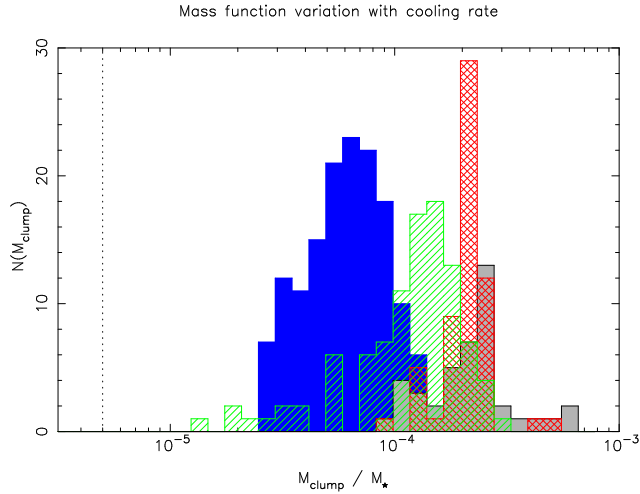}
        }
        \caption{Mass functions of clumps in simulations with different cooling times.  The mass functions were computed 0.7 local cooling times after the formation of the first fragment (i.e.~at 0.7 on the time axis in Fig.\ref{fig:cooling}).  The black/grey, red, green and blue MFs denote the simulations with $\tau=\infty$, 5, 3 and 2 respectively.  The dotted line denotes the approximate (mass) resolution limit of $10N_{\mathrm {ngb}}m$; all of our clumps are above this limit by at least a factor of 5--10.  We see clearly that faster cooling results in a smaller characteristic mass, and a somewhat broader mass function.}
        \label{fig:MFs}
\end{figure}

In Section \ref{sec:res_conv} we found that the MR runs were sufficient to resolve the fragmentation process properly across the full range of cooling times.  We now compare the MR runs across the full range of cooling times, in order to study the effect of the cooling rate on the fragmentation process.  The results of the MR runs are shown in Figs.\ref{fig:panels}--\ref{fig:MFs}.  Fig.\ref{fig:panels} shows the evolution of the disc surface density in the four different simulations, and it is clear that the cooling rate has a dramatic effect on the fragmentation process.  Slow cooling leads to the formation of  larger, more massive fragments with larger separations, while faster cooling results in a larger number of less massive fragments, with smaller separations.  Faster cooling also leads to the formation of more binary and multiple systems, although numerical dissipation results in many of these systems merging rapidly in our simulations.  Additionally, the fragmentation process appears to be more dynamic in the case of slow cooling, with individual objects showing greater perturbations from circular orbits.  This occurs because in the case of slow cooling the collapsing filament has time to become unstable to bending modes before it fragments, while with faster cooling the filament fragments before it becomes significantly bent.  

This qualitative assessment is seen more quantitatively in Fig.\ref{fig:cooling}, which shows the number and total mass of the bound fragments in each simulation.  Although mass is accreted into fragments more rapidly when the cooling is fast, the total mass in fragments does not differ significantly with different cooling times.  However, the simulations with faster cooling produce many more fragments, with correspondingly lower masses, and the incidence of multiple systems is greatly increased.  

Fig.\ref{fig:MFs} shows how the clump mass function (MF) varies as a function of the local cooling time.  We compare the MFs at a point where an approximately equal fraction of the disc mass is bound into clumps in each simulation, and before the (under-resolved) binary mergers begin to dominate the simulations with the fastest cooling.  We therefore compare the MFs at 0.7 local cooling times after the formation of the first fragment (see Fig.\ref{fig:cooling}), although we note that the MF comparison is qualitatively similar throughout the simulations.  All of the MFs are reasonably well-fit by log-normal functions, in agreement with previous studies of disc fragmentation \citep[e.g.][]{nayak07,rda08}.  We find that faster cooling results in a smaller characteristic fragment mass.  This is in agreement with the previous study of \citet{nayak07}, which also found that the characteristic fragment mass decreased as the cooling rate increased.  In our simulations the local (density-independent) cooling time when fragments form varies by a factor of $\simeq7$ between the slowest- ($\tau=\infty$) and fastest-cooling ($\tau=2$) models, and the characteristic mass of the respective MFs differs by a factor of $\simeq 4$.  We also find that the MFs become somewhat broader with shorter cooling times, but given the relatively small numbers of clumps formed in the $\tau=\infty$ and $\tau=5$ runs this result is not statistically significant.  The range in characteristic masses we find is much smaller than that seen by \citet{nayak07}, who found that the characteristic mass differed by a factor of $\simeq100$ between their simulations with $\beta=3$ and $\beta=0.3$.  We attribute this to the differences between the cooling laws adopted here and those used by \citet{nayak07}.  In particular, we note that the density dependence of our cooling law (which suppresses cooling at high densities) was not included in the simulations of \citet{nayak07}, and the likely consequence of this omission is to enhance fragmentation as the density in clumps increases.  We therefore suggest that the variation in characteristic mass with cooling time in real discs is less pronounced than that found by \citet{nayak07}, and conclude that large variations in the cooling time-scale result in only modest variations in the clump MF.

As mentioned above, a further difference between the fast and slow cooling simulations is in the fraction of multiple and binary fragments that are formed. The binary fraction is difficult to quantify due to the rapid merging of the binaries that form, but the differences are clearly seen in Fig.\ref{fig:panels}.  With critical cooling ($\tau=\infty$) most objects form as singles, and although some binaries form due to dynamical capture events the overall binary fraction is low.  However, when the cooling is faster the disc is thinner (colder) when it fragments, and consequently the clumps are more closely-spaced when they first form.  This leads to many more capture events, and we see many more binaries and multiples in the simulations with faster cooling.  Indeed, in the simulations with $\tau=3$ and $\tau=2$ almost all the clumps appear in binary or multiple systems.  We see no fragmentation of single clumps after the initial fragmentation occurs, and all of the binary and multiple systems observed form via dynamical capture events (in which the disc gas acts as a ``third body'').  We attribute the absence of any ``fission'' binaries to the density dependence of the cooling time, as the increase in $\beta$ with increasing clump density means that the value of $\beta$ in the high-density clumps rarely, if ever, falls below unity.  In this case (as explained in Section \ref{sec:sims}) we do not expect to see sub-clump fragmentation, and it seems that the formation of binaries by fission is unlikely in real discs with optically thick cooling.  The high incidence of capture binaries can easily be understood, as faster cooling leads to a lower temperature, and therefore a smaller disc scale-height $H$, in the fragmenting disc.  The length-scale that is most unstable to gravitational instabilities is $\simeq H$, so the typical fragment mass is $\simeq \pi \Sigma H^2$ and the typical fragment separation is $\simeq H$.  Faster cooling results in a smaller value of $H$, which in turn results in smaller fragment masses and smaller separations between fragments.  In our simulations these binaries tend to merge rapidly due to the high rate of numerical dissipation in their ``circumstellar'' discs (as seen between $t=5.1$ and $t=5.5$ in the $\tau=2$ simulation in Fig.\ref{fig:panels}), but in reality such mergers would likely be much less efficient.  Consequently, we find that fast cooling in fragmenting self-gravitating discs leads to a slight decrease in the characteristic fragment mass, and a dramatic increase in the incidence of binary and multiple objects.

%%%%%%%%%%%%%%%%%%%%%%%%%
\section{Discussion}\label{sec:dis}
Our simulations are robust and well-understood from a numerical perspective, but concerns remain when we consider the physical interpretation of our results.  The most obviously unphysical aspect of our simulations is our ``thin disc approximation'', using an artificial cooling law to allow fragmentation characteristic of a thin disc to be modelled at moderate numerical resolution.  It would be preferable to construct global models of thin discs, but this is not computationally feasible with current techniques, especially in the case of short cooling times.  Our disc model is effectively a local approximation, so the primary concern regards the boundary conditions.  In our models the real physical boundaries (at $r=1$ and $r=3$) are sufficiently far from the region of interest that they have no influence on the results, and the radial variation of the cooling time is negligible in the fragmentation region.  However, in a real thin disc there are many such filaments, and the typical radial separation between filaments can be as little as a few times $H$.  Dynamical interactions between adjacent filaments seem likely, especially once bound, point-mass-like objects have formed, and the likely outcome of this increased rate of clump-clump interactions is the formation of even more capture binaries than seen in our simulations.  In some of our simulations (those with slower cooling) some of the bound clumps undergo significant radial scattering even in the absence of adjacent filaments (moving as far as $\Delta r\simeq 0.3$ from their initial location), and consequently their cooling times change in an unphysical way.  However, this only occurs at late times in the simulations (to which we do not attach great significance), and moreover in such dense clumps the dominant term in the cooling time is the density dependence.  Consequently we are satisfied that our artificial cooling model is robust, and although our simulations should be regarded as numerical experiments rather than physical simulations, we are confident attaching limited physical significance to our results.  

An additional concern regards the density dependence of our cooling law.  Fixing the cooling time to scale as $\rho^{1/3}$ is correct only if the opacity is constant and the disc is optically thick, but it is not clear that these assumptions hold in reality.  In real self-gravitating discs the temperatures are expected to be $\sim100$K, so dust grains dominate the disc opacity \citep[e.g.][]{bl94,semenov03}.  In the case of a fragmenting disc with $Q=1$, the disc is optically thick if the orbital period is $\lesssim 1000$yr \citep{levin07}.  Real discs are generally only thought to be self-gravitating in their outer regions, at radii $\gtrsim50$AU in the case of protostellar discs \citep[e.g.][]{sw08} or at radii $\gtrsim0.1$pc in the case of a disc around a $\sim10^6$M$_{\odot}$ black hole \citep[e.g.][]{kp07}, and in both cases this suggests that fragmenting discs are marginally optically thick.   Individual clumps (with higher densities) thus have optical depths much greater than unity, so the optically thick approximation appears valid across a wide range of disc scales.  It is clear, however, that the opacity is not strictly constant in real discs.  Dust opacity is essentially independent of density, but varies strongly as a function of temperature: the opacity scales as $T^2$ for $T\lesssim160$K, and is approximately constant for $160$K$\lesssim T\lesssim1000$K \citep{bl94}.  It is reasonable to assume that the temperature will increase somewhat during collapse, leading to an increase in the dust opacity, so it seems likely that our adopted function $D(\rho)$ somewhat over-estimates the efficiency of radiative cooling.  Thus cooling is likely slower than predicted by our simulations in real fragmenting clumps.  Models of core-collapse during star formation predict only moderate increases in temperature, however, so the uncertainly due to the temperature dependence of dust opacity is likely small.  We note in passing that we also neglect further complexities of the collapse process, such as changes in gas composition with increasing temperature (due to dust sublimation, H$_2$ dissociation, etc.): such issues are well beyond the scope of this work.  

An additional complication is that is still not clear whether or not it is possible to form a disc with $\beta<1$.  A coherent disc structure cannot be assembled in the presence of such rapid cooling, as the growing disc would simply fragment on a dynamical time-scale.  Various authors have argued in favour of rapid cooling in self-gravitating discs around massive black holes \citep[e.g.][]{bl01,goodman03}, but these arguments generally rely on the assumption of roughly steady-state conditions in the disc, rather than considering disc formation.  \citet{rafikov07} argued that convective cooling results in an cooling time of $\beta \simeq 0.1$, but again does not consider disc formation in detail.  By contrast, \citet[][see also \citealt{levin07}]{nayak06} argues against the presence of rapid cooling in a fragmenting, star-forming disc around a massive black hole, instead arguing that the disc mass is accreted gradually.  The cooling time therefore decreases slowly as the disc mass is accumulated, so fragmentation occurs at $\beta\simeq\beta_{\mathrm {crit}}$.  If, however, the disc was formed in a rapid ``accretion event'', it seems that fragmentation could occur with $\beta<1$.  In addition the thermodynamics of the disc formation process are not fully understood, and one can conceive of a scenario where the disc is assembled with a longer cooling time and then begins to cool faster as it evolves \citep[e.g.][]{clarke07}.  One possible mechanism appeals to the ``opacity gap'' at $10^3$--$10^4$K: if a disc was assembled from warm, $\gtrsim 1500$K gas then the condensation of dust at $\sim1000$--1500K would result in a dramatic increase in the cooling rate as the temperature dropped, potentially resulting in a cooling time-scale shorter than the dynamical time-scale.  Our simple model is not able to address this issue in detail, but this problem provides fertile ground for future research.

Despite the fact that our simulations show good numerical convergence with increasing particle number, numerical resolution is still an issue that must be considered with care.  In particular, as noted in Section \ref{sec:res_conv}, the dynamical interactions between  different clumps are under-resolved even in our highest-resolution simulations.  This is a common feature of hydrodynamic simulations which seek to model large dynamic ranges in length and density, and in our simulations this results in unphysically rapid mergers of binary and multiple systems (as seen, for example, in Fig.\ref{fig:panels}).  We treat this problem by simply stopping our calculations before such under-resolved mergers begin to dominate the simulations, and attach only limited significance to our results beyond the point at which the first such mergers begin to happen.  [Note, however, that real mergers between fragments are expected to occur in such systems, and indeed have been invoked to explain the large discrepancy between the typical fragment mass and the observed stellar masses near the Galactic Center \citep{levin07}.]  A more sophisticated approach could, for example, make use of ``sink particles'' \citep[e.g.][]{bate95}, modified thermal physics \citep[e.g.][]{jappsen05}, or sub-resolution physics \citep[e.g.][]{nayak07} in order to follow the calculations further forward in time, but in our models (and in particular in the simulations with fast cooling) it is not clear that such approaches would yield more accurate results.  Here, a better approach may be to ``re-simulate'' a small region of our computational domain at much higher resolution in order to study the inter-clump dynamics in more detail, but this is beyond the scope of this paper.

\subsection{Application to the Galactic Centre}
We now consider the implications of our results for the formation of stars near the Galactic Centre (GC).  Many young, massive stars are known to exist close to the black hole at the centre of the Galaxy \citep{ghez98,genzel03,ghez05,paumard06}, and many of the stars at $\simeq0.1$pc are known to form a coherent ring or disc \citep[e.g.][]{genzel03}\footnote{Note that there is some controversy as to whether one or two discs of young stars exist at the GC \citep{paumard06,lu06}.  For the purposes of this discussion this is unimportant: the existence of one disc of young stars is sufficient to motivate our arguments, and they are not significantly altered by the existence of a second disc.}.  A popular theory for the formation of these stars is fragmentation of an accretion disc due to gravitational instabilities \citep[e.g.][]{lb03,gt04,nc05}, and recent hydrodynamic simulations have found that the properties of the stars formed in this manner are broadly consistent with the observed GC stellar disc \citep[e.g.][]{nayak07,rda08}.  However, to date most of these simulations have used greatly simplified thermal physics, adopting scale-free cooling and/or simplified equations of state, and given the critical role of cooling in the development of gravitational instabilities this treatment is less than ideal.  

Additionally, for numerical reasons most previous simulations of the formation of the GC stellar disc have made use of relatively slow cooling time-scales (with $\beta\simeq3$), although \citet{nayak07} did run a single model with $\beta=0.3$ (as discussed in Section \ref{sec:res_cool}).  However, analytic estimates of the cooling times in black hole discs generally find that the cooling time should be short compared to the dynamical time \citep[e.g.][]{goodman03,rafikov07} although, as discussed above, how such systems form remains uncertain.  Estimates of the cooling parameter in a fragmenting disc at the GC range from $\beta\sim0.1$ \citep{rafikov07} to $\beta=\beta_{\mathrm {crit}} \simeq 3$ \citep{nayak06}, which is approximately the range spanned by our simulations.  As discussed above, we find that a dramatic decrease in the cooling time leads to only a modest decrease in the characteristic stellar mass, and given the dependence of the characteristic mass on other disc properties (notably the disc aspect ratio $H/R$), this is unlikely to have a significant effect on the MF of the GC stellar disc.  However, our simulations indicate that fast cooling results in a very high primordial binary/multiple fraction (of order unity), and although binaries do form when cooling is slower, they are less common.  We therefore suggest that, if the GC stellar disc formed by fragmentation of a gaseous accretion disc, a high binary fraction may be indicative of rapid cooling in the parent accretion disc.

It is not clear, however, if this prediction provides a means of distinguishing between different proposed scenarios for the formation of the GC stellar disc.  The massive star population in the field is known to have a high multiplicity fraction, and the binary fraction for Wolf-Rayet stars in the solar neighbourhood is $\sim 40$\% \citep{vdhucht01}.  Little is currently known about binary statistics in the GC stellar cluster, with neither observations nor dynamical arguments \citep[e.g.][]{cuadra08} providing real constraints.  It is entirely possible (and in fact likely) that many unresolved close binaries near the GC may be masquerading as single objects in current data, but some binary systems have been observed.  In particular IRS 16SW, an eclipsing binary with a period of $\simeq$19.5 days, appears to be a contact binary consisting of two approximately equal-mass Wolf-Rayet stars with masses $\gtrsim 50$M$_{\odot}$ \citep{martins06,peeples07,rafelski07}.  Systems of this type appear to be a plausible outcome of fast cooling in a fragmenting, self-gravitating accretion disc, and we suggest that such systems should be common if the stellar disc at the GC formed via fragmentation of a gaseous accretion disc.

%%%%%%%%%%%%%%%%%%%%%%%%%%%
\section{Summary}\label{sec:summary}
We have used numerical hydrodynamic simulations to investigate the role of the cooling rate in the fragmentation of gravitationally unstable accretion discs.  Our simulations make use of a simple cooling law, designed to mimic the behaviour of radiative cooling in an optically thick, geometrically thin disc.  We performed detailed numerical convergence tests, and are satisfied that our simulations are well-resolved.  We find that decreasing the cooling time results in the formation of a larger number of fragments, but with a correspondingly smaller characteristic mass.  We also observed a high incidence of binaries, all of which form by dynamical capture.  Rapid cooling, on a time-scale shorter than the local dynamical time, results in a dramatic increase in the incidence of binary and multiple systems, although we are not able to determine the final properties of such systems.  We have considered our results in the context of the young stellar disc(s) at the Galactic Centre, and find that - at the very least - observations of some stellar binaries close to the Galactic Center do not rule out formation in a fragmenting accretion disc.

%%%%%%%%%%%%%%%%%%%%%%%%%%%
\section*{Acknowledgements}
We thank Yuri Levin, Sergei Nayakshin and Giuseppe Lodato for useful and stimulating discussions.  We also thank Volker Springel for providing us with some subroutines that are not part of the public version of {\sc gadget2}.  PJA thanks Andrea Ghez and the UCLA astronomy department for hospitality during the completion of this work.
This research was supported by NASA under grants NNG04GL01G and NNG05GI92G, by the NSF under grant AST 0407040, and in part by the NSF through TeraGrid resources provided by NCSA.
RDA acknowledges support from the Netherlands Organisation for Scientific Research (NWO) through VIDI grants 639.042.404 and 639.042.607.   

%%%%%%%%%%%%%%%%%%%%%%%%%%%

\label{lastpage}

\end{document}